\documentstyle[11pt,paspconf]{article}

\begin{document}
\title{Pixel and Micro-lensing with NGST}
\author{S. Bensammar\altaffilmark{1}}
\affil{Observatoire de Paris, France}
\author{A.-L. Melchior\altaffilmark{2}}
\affil{Queen Mary and Westfield College, London, UK}
\altaffiltext{1}{E-mail: bensammar@mesiob.obspm.fr} 
\altaffiltext{2}{Observatoire de Paris, France; E-mail:
A.L.Melchior@obspm.fr} 

\begin{abstract}
Within 8 years, the current microlensing surveys of M\,31 will provide
several hundred events affecting unresolved stars. They will thus
allow a statistical study of the dark matter in M\,31s halo. The NGST
will resolve these stars and constrain the mass of the corresponding
lenses. In case of on-line alerts from ground-based observations,
real-time NGST follow-up with high signal-to-noise ratio will provide
further constraints on the lenses. In addition, high resolution
observations with NGST will complement XMM and the previous optical
data and thus enable a closer insight of X-ray binaries within M\,31 to
be obtained. The optimal instrumentation to achieve these scientific
goals will be discussed. Last, the study of the dark matter
encompassed in the galaxy clusters would be possible with high angular
resolution observations on a large field camera and would open a new
field of research.
{\it Presented at NGST Science and Technology Exposition at Hyannis --
September 1999 -- (PASP) }  
\end{abstract}

\keywords{Microlensing, dark matter, infra-red, X-ray binaries}

\section{Introduction}
Searching microlensing events towards neighbouring galaxies allows to
probe the distribution of compact objects (also called MAssive Compact
Halo Objects or Machos) along their lines of sight (Paczynski 1986,
Griest 1991)\nocite{Paczynski:1986,Griest:1991}. The interpretation of
the first results is still very preliminary and relies on a small
number of events (Alcock et al.\ 1997a,b,c, Renault et al.\ 1997,
Pa\-lan\-que-Delabrouille et~al.\ 1998, Alard et al.\ 1997, Udalski et
al.\ 1994, Derue et al.\
1998)\nocite{Alcock:1997d,Alard:1997,Udalski:1994b}\nocite{Derue:1998}\nocite{Alcock:1997a,Alcock:1997b,Renault:1997,Palanque-Delabrouille:1998}.
Enlarging this faint statistics and exploring new lines of sight will
be key-points in the near future for drawing strong constraints on the
dark haloes. The detection of microlensing events on unresolved stars,
initiated by Crotts (1992) and Baillon et al.\
(1993)\nocite{Crotts:1992,Baillon:1993}, offers some new
perspectives. Two independent methods, based on this principle, have
been developed. On the one hand, monitoring the fluxes of all the
pixels present on the images allows to achieve a good sensitivity to
the possible variation of unresolved stars (Ansari et al., 1997;
Melchior et al.\
1998,1999\nocite{Ansari:1997,Melchior:1998,Melchior:1999}).  A
systematic study of all the information present in the frames is thus
possible, with in particular an estimation of the detection
efficiencies.  On the other hand, the image subtraction technique
also allows the detection of variable objects (Crotts \& Tomaney
1996\nocite{Crotts:1996}), but has mainly been used so far to improve
the photometry of microlensing events detected towards the Galactic
Bulge and LMC (e.g. Alard 1999; Alcock et al.\
1999\nocite{Alard:1999b,Alcock:1999}).

Thus, when accounting for unresolved stars, the number of effectively
monitored stars is significantly enlarged, lines of sight towards more
distant target galaxies can be explored, and hence the number of
potential target galaxies is increased by an order of magnitude. The
efficacy of these approaches to detect luminosity variations with a
high detection rate has been demonstrated. The flux of the unresolved
star is by definition unknown: this prevents the definition of the
Einstein ring crossing time (=intrinsic duration of the events) and
the estimation of the lens mass.  In this context, the NGST will open
new opportunities. Here, we discuss a few of them with respect to the
possible instrumentation of this telescope.

\section{Towards M\,31}
The survey of M\,31 undertaken at the Isaac Newton
Telescope\footnotemark{3}\footnotetext[3]{see
http://www.ast.cam.ac.uk/$\sim$mike/casu/WFCsur/M\,31.html and\\
http://www$-$star.qmw.ac.uk/AGAPE/} will detect a few hundred
microlensing events\footnotemark{4}\footnotetext[4]{This typically
assumes a 5-year monitoring of M\,31.} if the dark haloes of M\,31 and
the Milky Way are filled with compact objects.  In the following, we
discuss how the NGST could improve our knowledge of the lenses and the
dark component of the haloes.
\subsection{Identification of unresolved stars for microlensing
(Integral Field Spectrometer with R$\sim$1000)} Whereas spectroscopic
identifications of unlensed (resolved) stars in the Galactic Bulge
(e.g. Benetti et al. (1995)) are possible from the ground, the NGST
will offer the possibility to perform a similar work in M\,31 with
unresolved stars.  With a small field of view (10''$\times$10'') in
the optical and the near-infrared, the spectroscopy (R$\sim$1000) with
high spatial resolution (0.1'') would allow to further study
microlensing candidates detected from ground-based telescopes, and in
particular those affecting unresolved stars in M\,31. The
identification of those unresolved sources is very challenging with
existing technology. For instance, the microlensing candidate detected
by the AGAPE group (Ansari et al.\ 1999)\nocite{Ansari:1999}, at 41''
from M\,31's centre, affects a star whose magnitude at rest is dimmer
than 22 in R, and lies on a stellar background of magnitude 16
mag.arcsec$^{-1}$. In this case, 3D imaging from space will be
possible and allows an identification and detailed study of the
(unlensed) star.

The unresolved stars corresponding to the microlensing events detected
during the on-going ground-based surveys towards M\,31 (e.g. INT
survey) could be resolved and further studied with 3D imaging. This
would provide complementary informations and place some constraints on
the lens mass. The typical expected events will occur in an area with
a surface magnitude of 22, and the NGST spectroscopy would easily
detect stars down to magnitude 26. Moreover, spectroscopy in the
near-infrared will be possible for the study of red giants.
 
\subsection{Identification of unresolved stars for Low Mass X-ray Binaries
(Integral Field Spectrometer with R$\sim$200)} These microlensing
surveys will also achieve a good sensitivity to cataclysmic variables.
XMM\footnotemark{5}\footnotetext[5]{see
http://astro.estec.esa.nl/XMM/xmm$\_$top.html,
http://xmmssc-www.star.le.ac.uk/ and\\
http://www-star.qmw.ac.uk/AGAPE/xpage1.html}, due to be launched in 15
December 1999, will observe M\,31. The cross-identification with the
microlensing surveys will provide the first optical counter-parts of
LMXB. The NGST will be complementary to these two observing modes, and
in particular, be able to identify the optical counter-parts of LMXB
{\em when quiescent} (mag$\sim$28-30). In fact, the optical to
near-infrared spectrum could be studied, and thus allow an
unprecedented study of these systems in M\,31.

\subsection{Real-time follow-up of ground-based observations (Integral Field 
 Spectrometer with R$\sim$200-2000)} Whereas spectroscopy of on-going
microlensing events towards the Galactic bulge has been performed
(e.g. see the remarkable work of Lennon et al.\
(1996)\nocite{Benetti:1995,Lennon:1996}), a similar work, but for
extragalactic stars, will be possible with the NGST.  If ground-based
microlensing surveys towards galaxies like M\,31 are still going-on in
2007, high S/N follow-up observations based on an alert system would
be possible with 3D imaging. Possible deviations from the point
source/point lens approximations could be studied in external galaxies
like M\,31: (1) chromatic and spectroscopic signatures as suggested by
Valls-Gabaud (1998) (and references therein) for the study of stellar
structure; (2) planets orbiting around a star in M\,31 (e.g. di Stefano
1999\nocite{diStefano:1999}); (3) binary lenses (e.g. Gaudi \& Gould
1999)\nocite{Gaudi:1999}.

\subsection{New survey of M\,31 in the near-infrared (Wide Field Imager --
multi-band filter mode)} Depending on the results of the current
ground-based surveys, monitoring a large field of view (5'$\times$5')
in M\,31 with a resolution better than 0.03'' would allow a
complementary study. A sensitivity to different (dimmer) stars could
be achieved, and in addition to the optical, the near-infrared
wavelength range (1-2$\mu$m) could also be investigated. Also the
constant seeing, which would characterise such data, would be a major
advantage.  A multi-band follow-up will allow an unprecedented
analysis of microlensing events, with in particular a determination of
the lens mass, completely independent from the ground-based optical
observations. Moreover, a close insight into extragalactic dim stars
will be possible for the first time.

\section{Towards other galaxies}
A near-infrared survey, with high spatial resolution (0.03''),
combined with a large field of view (5'$\times$5'), will allow to
probe more distant galaxies, and thus a larger number of lines of
sight can be explored. In particular, M\,87 (at $\simeq$ 17~Mpc) will be
seen from space (with a constant seeing $\simeq 0.03$'') with the same
observing conditions as M\,31 (at 0.7~Mpc) with ground-based
observations (with a mean seeing $\simeq 1.5$'').  A space-based
monitoring of galaxies at the distance of M\,87 with an 8-meter
telescope will require 5-min exposures. In first approximation, if
M\,31 was at the same distance as M\,87, we could typically expect of
order of 50 microlensing events over a 6 months observing period. As
studied by Gould (1995)\nocite{Gould:1995b}, additional events could
be expected if intra-cluster MACHOs constitute a significant fraction
of the dark matter of the Virgo cluster.  An ambitious programme
monitoring 10-20 galaxies would perform an unprecedented cartography
of dark haloes.

Real-time follow-up of such microlensing events with 10~mas resolution
would achieve a high signal-to-noise ratio and to constrain the lens
mass.

\section{Conclusions}
The NGST equipped with an Integral Field Spectrometer
(R$\sim$200-2000) and with a multi-band filter mode on a Wide Field
Imager, characterised by a good spatial sampling, would first allow
for the first time the study of extragalactic dim stars thanks to the
magnification detected with ground-based observations. With a wide
spectral range (from optical to near-infrared), the NGST would be
unique to break the degeneracy of the parameters of each lens-source
system as it could resolve the stars. Hence, the mass function of the
lenses detected in M\,31 could be determined, and a better understanding
of the dark matter content of dark haloes achieved. Exotic events in
M\,31 could be studied with an unprecedented sensitivity and would
further help to constrain the lens population. In addition, this would
offer the unique opportunity to detection of extragalactic planets
(see Rhie et al. (1999)\nocite{Rhie:1999} for the possible detection
of the first planet in the Galactic Bulge).

\acknowledgments 
During this work, A.-L. Melchior has been supported by a European
 contract ERBFMBICT972375 at QMW.


\begin{references}
\reference Alard C.: 1999, \newblock { A$\&$A}, { 343}, 10

\reference Alard C., Guibert J.: 1997, \newblock { A$\&$A}, { 326}, 1 

\reference Alcock C. et al., 1997a, \newblock \apj, { 486}, 697

\reference Alcock C. et al., 1997b, \newblock \apj, { 491}, L11

\reference Alcock C. et al., 1997c, \newblock \apj, { 479}, 479

\reference Alcock C. et al., 1997d, \newblock \apj, { 521}, 602

\reference Ansari R. et al.,   1997, \newblock { A$\&$A}, { 324}, 843

\reference Ansari R. et al. 1999, \newblock { A$\&$A}, { 344}, L49

\reference Baillon P. et al., 1993, \newblock { A$\&$A}, { 277}, 1

\reference Benetti S., Pasquini L., West R.~M., 1995, \newblock { A$\&$A},
{ 294}, L37

\reference Crotts, A. P.~S., 1992, \newblock \apj, { 399}, L43

\reference Crotts A. P.~S., Tomaney A.~B., 1996, \newblock \apj, { 473},
L87

\reference Derue F. et~al., 1998, \newblock { Prelimnary results of
the EROS 2 experiment towards the Spiral Arms of the Galaxy},
\newblock 4th International Workshop on Gravitational Microlensing
Surveys - Paris

\reference  di~Stefano R., Scalzo R.~A., 1999,
\newblock \apj, { 512}, 564

\reference Gaudi
B.~S., Gould A., 1999, \newblock \apj, { 513}, 619

\reference Gould A., 1995, \newblock \apj, { 455}, 44G

\reference Griest K., 1991, \newblock \apj, { 366}, 412

\reference Lennon D.~J. et al., 1996, \newblock \apj, { 471},
L23

\reference Melchior, A.-L. et~al.: 1998, \newblock { A$\&$A}, { 134}, 377

\reference Melchior, A.-L. et~al.: 1999, \newblock { A$\&$AS}, { 339}, 658

\reference Paczynski B., 1986, \newblock \apj, { 304}, 1

\reference Palanque-Delabrouille
N. et~al., 1998, \newblock { A$\&$A}, { 332}, 1

\reference Renault C. et al., 1997, \newblock { A$\&$A}, { 324}, L69

\reference Rhie S.~H. et al., 1999, \newblock { astro-ph/9905151}

\reference Udalski A. et al., 1994, \newblock { Acta Astron.}, { 44}, 165

\end{references}
\end{document}